\documentclass{article}

\usepackage[nonatbib,preprint]{neurips_2022}

\usepackage[utf8]{inputenc} 
\usepackage[T1]{fontenc}    
\usepackage{hyperref}       
\usepackage{url}            
\usepackage{booktabs}       
\usepackage{amsfonts}       
\usepackage{nicefrac}       
\usepackage{microtype}      
\usepackage{xcolor}         
\usepackage{tabto}    
\usepackage[superscript,biblabel]{cite}
\usepackage{setspace}  
\usepackage{titlesec}
\usepackage{appendix}
\usepackage{multirow}
\usepackage[ruled]{algorithm2e}
\usepackage{subfig}
\usepackage{wrapfig}
\usepackage{tikz}
\usepackage{tikz-cd}
\usetikzlibrary{cd}
\usepackage{microtype}
\usepackage{graphicx}
\usepackage{booktabs} 
\usepackage{pifont}       
\usepackage{bbding}       %
\usepackage{fontawesome}  
\usepackage{jabbrv} 

\usepackage{amsmath}
\usepackage{amssymb}
\usepackage{mathtools}
\usepackage{amsthm}
\usepackage{arydshln}
\usepackage[capitalize,noabbrev]{cleveref}

\makeatletter
\renewcommand{\fnum@figure}{\textbf{Fig. \thefigure} | }
\renewcommand{\fnum@table}{\textbf{Table \thetable \ |}}

\makeatother
\usepackage[labelsep=space]{caption}
\usepackage[
singlelinecheck=false 
]{caption}

\usepackage{authblk}

\title{
Lightening Anything in Medical Images
}

\author[1,*]{Ben Fei}
\author[1,*]{Yixuan Li}
\author[1,$\dagger$]{Weidong Yang}
\author[3,$\dagger$]{Hengjun Gao}
\author[1]{Jingyi Xu}
\author[1]{Lipeng Ma}
\author[4]{Yatian Yang}
\author[2]{Pinghong Zhou}
\affil[1]{School of Computer Science, Fudan University, Shanghai, China, 200438}
\affil[2]{Endoscopy Center of Zhongshan Hospital, Fudan University, Shanghai, China, 200032}
\affil[3]{Institute of Digestive Disease, School of Medicine, Tongji University, Shanghai, China, 200065}
\affil[4]{National Engineering Center for Biochip, Shanghai, China, 201203}
\affil[$\dagger$]{Corresponding to: wdyang@fudan.edu.cn, hengjun\_gao@tongji.edu.cn}

\begin{document}

\maketitle

\begin{abstract}
The development of medical imaging techniques has made a significant contribution to clinical decision-making. However, the existence of suboptimal imaging quality, as indicated by irregular illumination or imbalanced intensity, presents significant obstacles in automating disease screening, analysis, and diagnosis. Existing approaches for natural image enhancement are mostly trained with numerous paired images, presenting challenges in data collection and training costs, all while lacking the ability to generalize effectively. Here, we introduce a pioneering training-free Diffusion Model for Universal Medical Image Enhancement, named UniMIE. UniMIE demonstrates its unsupervised enhancement capabilities across various medical image modalities without the need for any fine-tuning. It accomplishes this by relying solely on a single pre-trained model from ImageNet. We conduct a comprehensive evaluation on 13 imaging modalities and over 15 medical types, demonstrating better qualities, robustness, and accuracy than other modality-specific and data-inefficient models. By delivering high-quality enhancement and corresponding accuracy downstream tasks across a wide range of tasks, UniMIE exhibits considerable potential to accelerate the advancement of diagnostic tools and customized treatment plans.
\end{abstract}

\setstretch{1.8}

\section{Introduction}
\label{sec:introduction}
\begin{figure}[t]
    \centering
    \includegraphics[width=\linewidth]{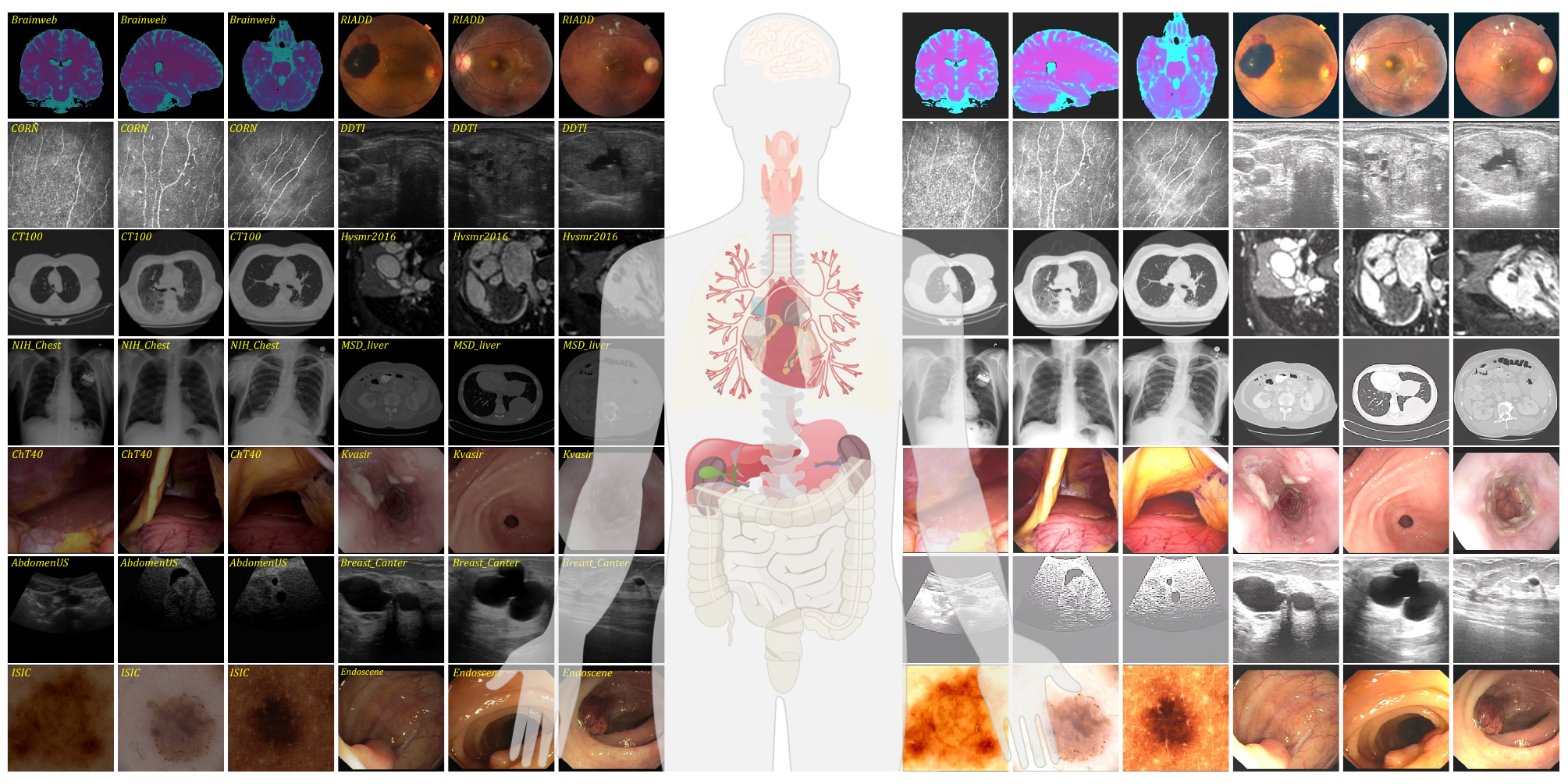}
    \caption{\textbf{UniMIE is pre-trained on ImageNet but can handle diverse enhancement tasks.} The generative prior inherent in UniMIE can be generalized to any medical image modality even if it is not in the training set.}
    \label{fig:teaser}
\end{figure}

The field of clinical medicine has undergone a revolutionary transformation thanks to the rapid advancements in medical imaging technology. Medical images have become invaluable resources for clinicians, offering a wealth of information pertaining to biological and anatomical tissues. Consequently, these images play a pivotal role in facilitating accurate diagnosis and effective treatment strategies.
Nevertheless, medical images exhibit considerable variations in quality, regardless of whether they are obtained using the same or different devices.
These variations can manifest as defects including low contrast, intensity inhomogeneity, inevitable blur, and noise, all of these can arise during the image acquisition process.
As depicted in Fig.~\ref{fig:teaser}, we present an illustrative example showcasing both low- and high-quality medical images. These medical images were captured through various imaging techniques such as confocal microscopy, color fundus cameras, endoscopy, and others.
Regarding the high-quality medical images (depicted in Fig.~\ref{fig:teaser}(\textbf{Right})), clinicians can readily discern and identify nearly all details. 
Conversely, the low-quality medical images (shown in Fig.~\ref{fig:teaser}(\textbf{Left})) pose challenges in clearly visualizing the complete structure of corneal nerve fibers, digestive tract, or other pertinent tissues and lesions.
When comparing medical images to natural images, it is important to note that medical images are typically acquired using specialized imaging techniques. These techniques can introduce unique degradation factors that are not commonly found in natural images.
As a result, these factors can cause various appearance artifacts that compromise the quality of the images, leading to additional challenges in the realm of clinical applications.
It is revealed that approximately 12 $\%$ of the fundus images, obtained from a sample of 5,575 consecutive patients, were considered unreadable by ophthalmologists due to inadequate quality~\cite{philip2005impact}. 
Similarly, the UK BioBank dataset, which includes approximately 30 $\%$ of retinal images, failed to meet the necessary quality standards for accurate diagnosis~\cite{welikala2017automated}.
Moreover, these limitations impede the effectiveness of subsequent tasks in image analysis, such as segmenting specific structures, detecting lesions, and facilitating computer-aided diagnosis.
Hence, the development of fully automated and reliable techniques for enhancing medical images has always been acknowledged as a crucial preliminary stage in clinical applications. 
These technologies are vital in producing high-quality medical images with fine-grained details, optimal brightness, and appropriate contrast.

Over the past few decades, various conventional methods have been designed for enhancing natural images, including histogram equalization (HE)~\cite{cheng2004simple}, dark channel prior (DCP)~\cite{he2010single}, as well as Retinex-based~\cite{rahman2004retinex} and filtering-based~\cite{ko1991center} techniques.
However, these methods often demonstrate sensitivity towards a restricted set of parameters, lack flexibility, and typically necessitate manual adjustment.
The predominant approaches for enhancing natural images based on deep learning heavily rely on fully supervised learning, which mandates the availability of aligned image pairs during the training process. 
Obtaining such pairs of low and high-quality medical images in real-world situations for training purposes poses significant challenges. 
Therefore, a limited number of unsupervised learning methods have been designed to tackle this predicament~\cite{chen2018deep, zhu2017unpaired, zhang2018multi, jiang2021enlightengan}. 
Nonetheless, these frameworks often exhibit instability, occasionally exacerbate noise, and may suffer from halo artifacts.

There has been a recent surge of interest in exploring more generalized image priors using generative models~\cite{shaham2019singan,gu2020image,asim2020blind} to effectively handle image restoration in an unsupervised manner~\cite{chen2018image,el2022bigprior}. 
In this particular context, the inference process allows for the simultaneous handling of multiple restoration tasks involving various degradation models, eliminating the requirement for fine-tuning.
For example, the utilization of Generative Adversarial Networks (GANs)~\cite{goodfellow2020generative}, trained on comprehensive collections of clean data, has proven effective in tackling a range of linear inverse problems via GAN inversion~\cite{pan2021exploiting,menon2020pulse,gu2020image}. Simultaneously, Denoising Diffusion Probabilistic Models (DDPMs) have demonstrated remarkable generative capabilities, providing detailed outputs and diverse outcomes when compared with GANs~\cite{ho2020denoising,sohl2015deep,song2020improved,ramesh2021zero}.

Consistent with these observations, we introduce UniMIE, a training-free Diffusion Model designed to enhance medical images without the need for fine-tuning on medical data. UniMIE accomplishes this by utilizing a single model that is pre-trained on ImageNet~\cite{dhariwal2021diffusion}.
UniMIE is comprehensively evaluated through a series of rigorous experiments conducted on 15 validation datasets (see Supplementary Table 1 and Fig. 1-15), covering a diverse range of medical imaging modalities, pathological conditions, and anatomical structures. 
The experimental results consistently demonstrate that UniMIE outperforms off-the-shelf image enhancement models, while achieving comparable or even superior results to specialist models specifically trained on the paired images. These findings underscore the potential of UniMIE as a transformative approach for versatile medical image enhancement.

\begin{figure}[t]
    \centering
    \includegraphics[width=\linewidth]{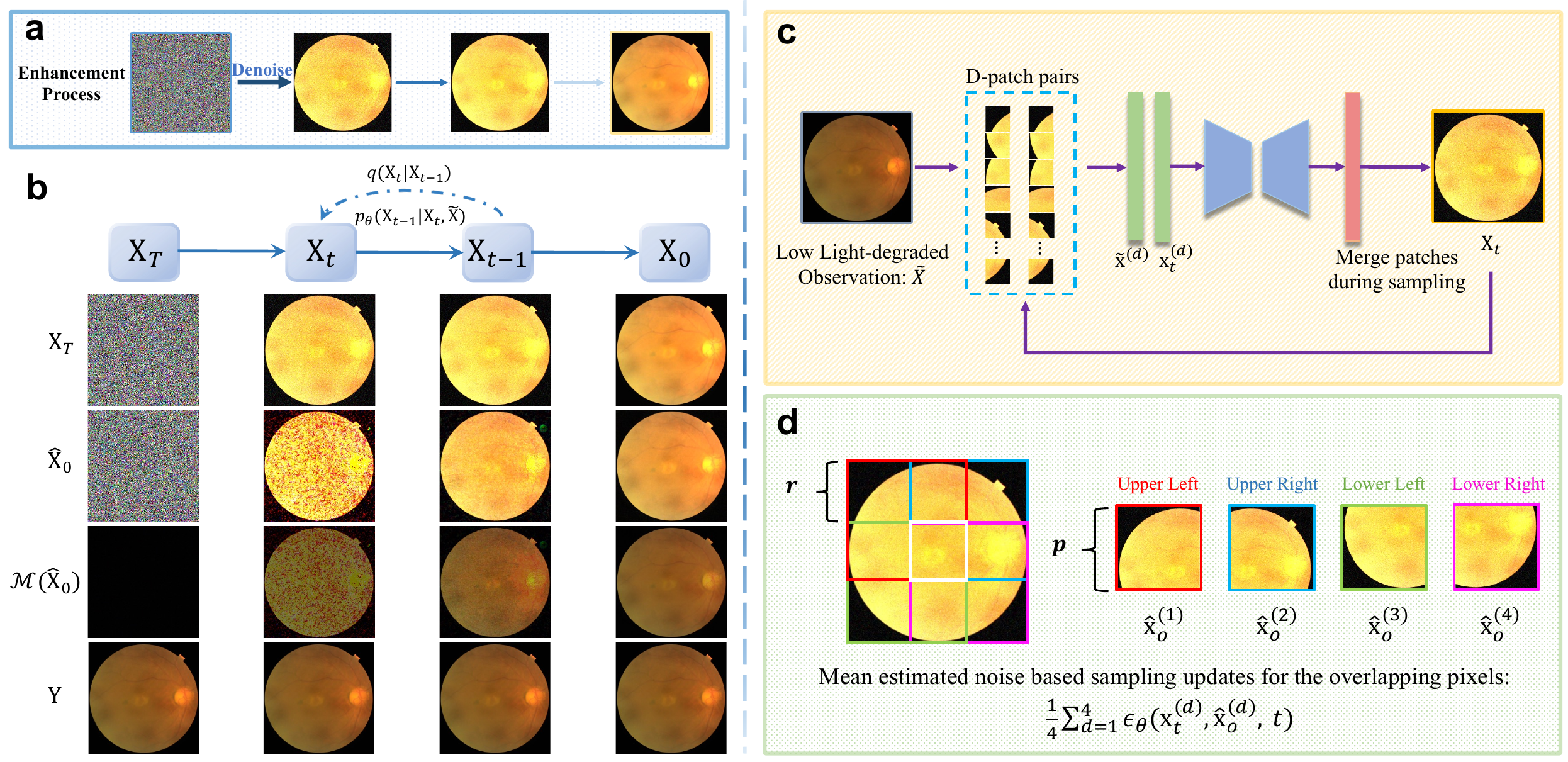}
    \caption{\textbf{Principle of UniMIE.} \textbf{a,} The enhancement process of our UniMIE. 
    \textbf{b,} An illustration of the forward process and the reverse process in an unconditional diffusion model. \textbf{c,} An overview of the patch-based pipeline utilized for medical image enhancement. \textbf{d,} An illustration of the mean estimated noise-guided sampling updates for overlapping pixels across patches, where the ratio of patch size to the overlapping region is $r = p/2$. Specifically, the grid cell highlighted with a white border is shared by only four overlapping patches. Consequently, sampling updates are performed for the pixels within this region at each time step $t$. These updates consider the average estimated noise across the four overlapping patches.}
    \label{fig:fig1}
\end{figure}

\begin{figure}[t]
    \centering
    \includegraphics[width=\linewidth]{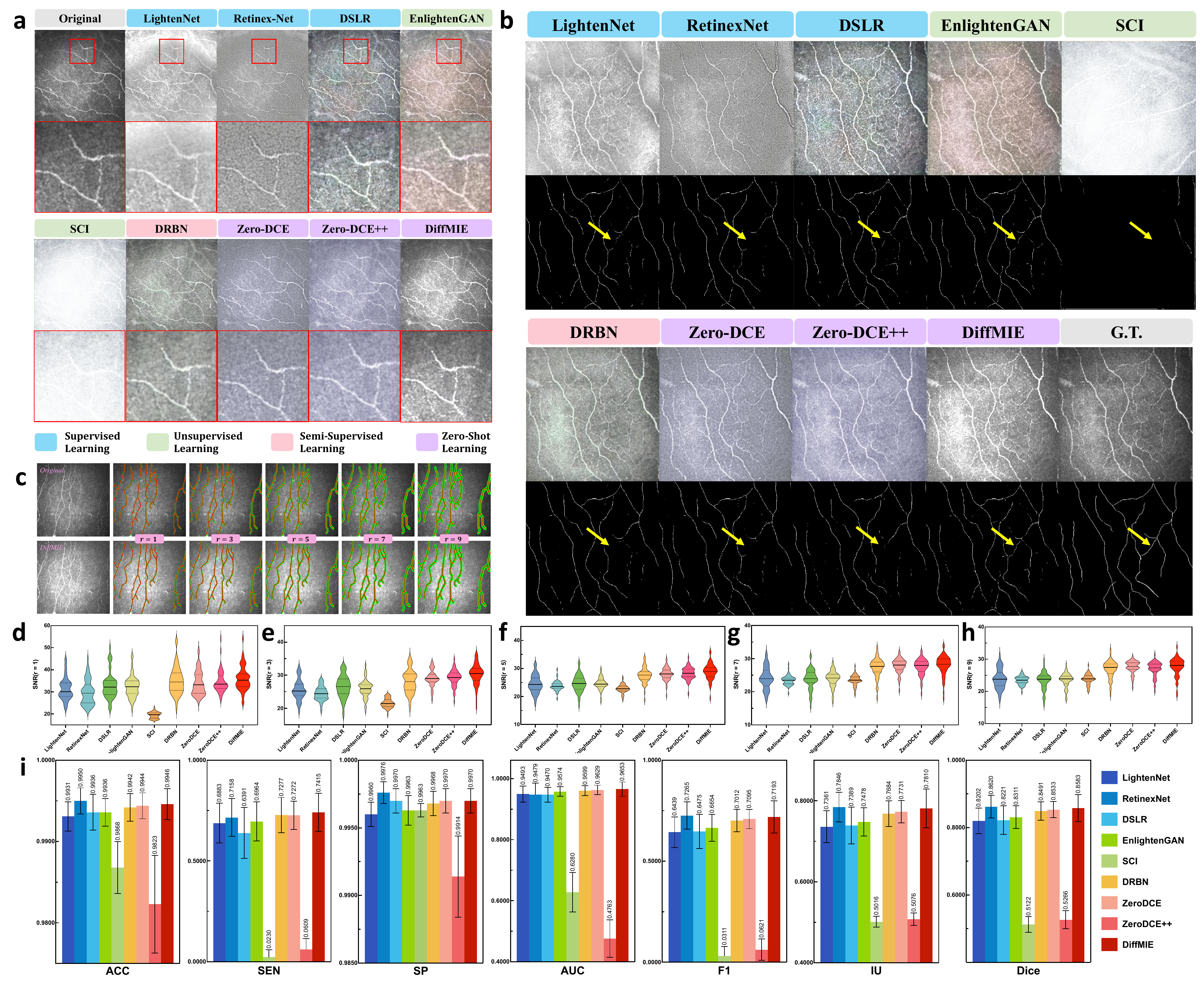}
    \caption{\textbf{a,} Comparison of noise artifact of enhanced corneal confocal microscopy images from CORN dataset. \textbf{b,} An illustrative example to demonstrate the selection of background regions for calculating the SNR. The background regions, depicted in green, were identified by applying a disk-shaped dilation manipulation on the manually traced fibers, which are represented in red. The dilation operation was performed using radii of 1, 3, 5, 7, and 9 pixels. The top row displays the low-quality image, while the bottom row showcases the image enhanced by UniMIE. \textbf{c,} An exemplification of enhanced CCM medical images. and corresponding nerve fiber segmentation performance. \textbf{d-h,} SNR results of the low-quality and enhanced CCM medical images by various methodologies. \textbf{i,} Segmentation performance of enhanced CCM medical images using various methods in terms of ACC, SEN, SP, AUC, F1, IU, and DICE.}
    \label{fig:fig2}
\end{figure}

\section{Results}
\label{sec:exper}

\subsection{Principle of UniMIE: A training-free framework for any medical image enhancement with any size}

The primary aim of our paper is to introduce a training-free model that fulfills the role of universal medical image enhancement. A key aspect in the development of such a model is its ability to adapt to a diverse array of variations in imaging conditions, anatomical structures, and pathological conditions.
To this end, we propose a universal and training-free pipeline of UniMIE shown in Fig.~\ref{fig:fig1}a.
UniMIE aims to utilize a single diffusion model ~\cite{dhariwal2021diffusion} pre-trained on ImageNet to solve any medical image enhancement with any size by regarding this problem as an inversion problem.
Specifically, during the sampling process of the unconditional diffusion model, each image $\mathbf{X}_{t}$ is utilized in every denoising step to estimate a temporary variable $\hat{\mathbf{X}}_{0}$ (Methods).
$\hat{\mathbf{X}}_{0}$ goes through a learnable degradation mask $\mathcal{M}$ to obtain $\mathcal{M}(\hat{\mathbf{X}}_{0})$.
The distance metric is applied between $\mathcal{M}(\hat{\mathbf{X}}_{0})$ and a provided low-quality medical image $\mathbf{Y}$.
The utilization of the gradient of this distance metric will guide the image sampling process in the next time step and enable the optimization of the learnable degradation mask.
A detailed implementation of sampling is depicted in Fig.~\ref{fig:fig1}b. 
After the enhancement process, the generated images obtain a medical level of quality based on the powerful prior inherent in unconditional diffusion models.

The diverse sizes of medical images obtained from different medical testing equipment also pose a challenge as deep learning models typically require images with fixed resolutions.
Compared with previous methods, UniMIE can not only accept images from different modalities but also medical images of any size.
The specific implementation route is shown in Fig~\ref{fig:fig2}c and d. The given degraded image will be segmented according to the step size $r$ of the sliding window.
The same operation will also be applied to each noisy image in the sampling process. 
In this way, during each step of the diffusion model's sampling process, conditional guidance is utilized to generate each patch of the image based on the corresponding patch in the degraded image.
More importantly, the interaction of the overlap between each patch makes the final image free of artifacts between patches.

\subsection{Confocal images enhancement and neural fiber segmentation}
Confocal microscopy images are often limited by factors such as noise, scattering, and quenching, which may result in reduced image quality or loss of information~\cite{wu2021multiview}. Confocal microscopy image enhancement technology can improve the contrast, resolution, and clarity of images to better demonstrate the details of cells and tissues~\cite{chen2023artificial}. This helps physicians more accurately observe and analyze the morphology, structure, and function of nerve cells and fibers.

We firstly examine the effectiveness of UniMIE for enhancing confocal images on the CORN-2 (CORneal Nerve) dataset~\cite{mou2019cs}. 
This dataset was created with the goal of enhancing confocal images, specifically those obtained from the publicly available Cornea Confocal Microscopy (CCM) dataset~\cite{mou2019cs}.
The original low-quality in this dataset exhibits non-uniform intensity and contains corneal scars or imaging speckles (Fig.~\ref{fig:fig2}a). 
Therefore, precisely distinguishing between nerve fiber structures and corneal scars or speckles is crucial for the effective enhancement of CCM medical images. 
We compare the enhancement performance of UniMIE with other commonly used natural image enhancement methods.
It is evident upon visual inspection that there exists a noticeable discrepancy in the presence of noise artifacts between the low-quality confocal image and the confocal image that has been enhanced utilizing different image enhancement techniques (Fig.~\ref{fig:fig2}a).
In contrast, UniMIE achieves more accurate enhancement of nerve fibers, thereby assisting healthcare professionals in clinical diagnosis (Fig.~\ref{fig:fig2}a and Supplementary Fig. 1).
Moreover, compared with other supervised~\cite{li2018lightennet,liu2021retinex,lim2020dslr}, unsupervised~\cite{jiang2021enlightengan, ma2022toward}, semi-supervised~\cite{yang2021band}, and zero-shot~\cite{guo2020zero,zhang2019zero}, the images enhanced by UniMIE significantly maintain the details of images.

To quantitatively evaluate the enhanced confocal images, we employ the Signal-to-Noise Ratio (SNR) metric to evaluate the manually annotated neural fibers~\cite{ma2021structure}. 
In the experimental setup,  the background area is precisely defined as the image region acquired through the application of a dilation operation in the shape of a disk, while excluding the signal region. To calculate SNR values, we apply dilation operations with radii of 1, 3, 5, 7, and 9 pixels, respectively.
Fig.~\ref{fig:fig2}c illustrates an example showcasing a signal region and background regions of varying radii. 
As observed in Fig.~\ref{fig:fig2}d-h, UniMIE surpasses all other natural image enhancement methods in terms of achieving the highest SNR.
These observations emphasize the efficacy of UniMIE in mitigating the issue of inconsistent lighting in the background regions of color confocal images, while effectively enhancing the visualization of signal areas.

To evaluate the influence of the proposed UniMIE on downstream tasks, we additionally conduct corneal nerve fiber segmentation on the corneal confocal images. We train the MMDC~\cite{zhong2022you} using the CORN-1 dataset and subsequently evaluate its performance on the enhanced corneal confocal images.
The qualitative segmentation results are compared in Fig~\ref{fig:fig2}b. It is observed that very subtle corneal nerve fibers can also be well-segmented on the enhanced images by UniMIE.
To comprehensively assess the performance of segmentation, we compute several metrics between the predicted and ground truth results, including Accuracy (ACC), Sensitivity (SEN), Specificity (SP), Area Under the Curve (AUC), F-score@$\%$ (F1), Intersection over Union (IU) as well as Dice coefficient (Dice), 
Without pre-training on any low- and high-quality medical image pairs, UniMIE outperforms other zero-shot methods in all metrics. 
Compared with off-the-shelf approaches, UniMIE fulfills the best SEN and AUC and the second-best ACC, SP, F1, IU, and Dice.

\begin{figure}[t]
    \centering
    \includegraphics[width=0.95\linewidth]{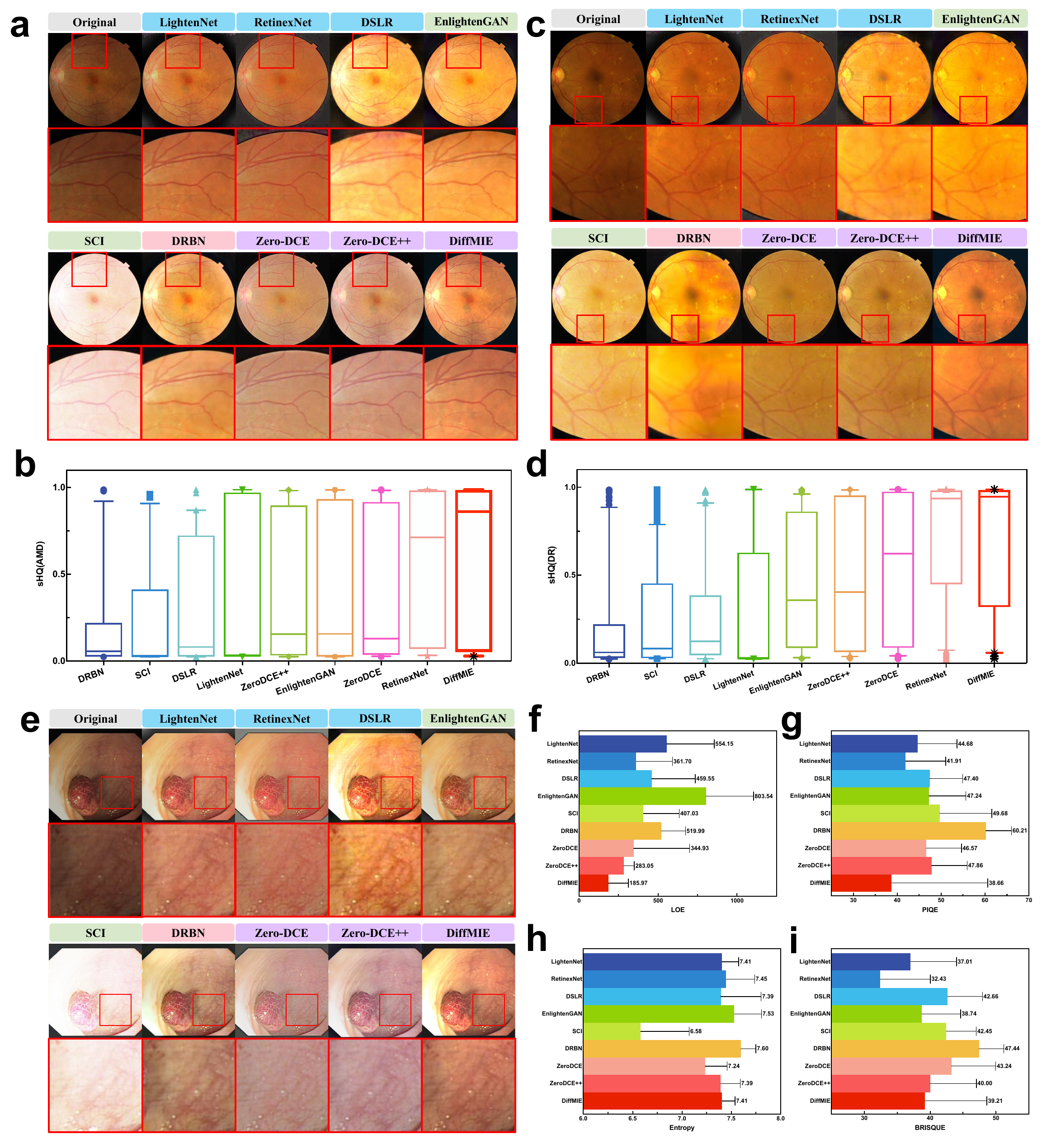}
    \caption{Comparison of texture details of enhanced color fundus images: \textbf{a,} DR and \textbf{b,} AMD eye diseases. High-quality score performance (mean $\pm$ standard deviation) of different enhancement methods on enhanced color fundus images: \textbf{c,} DR and \textbf{d,} AMD eye diseases. \textbf{e,} A qualitative examination is performed to compare the distribution of illumination in enhanced Endoscene images. \textbf{f-i,} Comparison of image quality metrics for different enhancement methods.}
    \label{fig:fig3}
\end{figure}

\subsection{Performance on color fundus images}

Color fundus images are an important tool in the evaluation of eye diseases, including retinal diseases, glaucoma, and macular degeneration~\cite{mitani2020detection}. By enhancing the image, the physician's visibility of lesions and diagnostic accuracy can be improved. The enhanced images can highlight blood vessels, lesions and other important anatomical structures, helping doctors better detect and identify signs of disease~\cite{poplin2018prediction}.

Therefore, we aim to provide additional validation for the efficacy of UniMIE in analyzing color fundus images, encompassing prevalent ocular conditions like diabetic retinopathy (DR) and age-related macular degeneration (AMD) sourced from RIADD dataset~\cite{pachade2021retinal}.
RIADD is part of the RFMiD dataset, publicly released by the World Health Organization in 2019.
The RIADD dataset exhibits significant variations in image quality, encompassing instances of under/over exposure, blurring, noise, and artifacts.
The disparities in texture details between the original low-quality medical images and the medical images that have been enhanced using various natural image enhancement methods are illustrated in Fig.~\ref{fig:fig3}a and c, Supplementary Fig. 2 and 3.
Notably, the qualitative results of enhanced color fundus images by UniMIE surpass those of supervised and self-supervised learning methods. 
We further conducted a quantitative analysis of UniMIE by calculating the enhanced retinal image quality scores. 
We employed a cutting-edge pre-trained classification model called MCF-Net~\cite{fu2019evaluation} to obtain the quality scores of the enhanced fundus images from two common datasets of eye diseases: DR and AMD. 
The quality score evaluates the overall perceptual quality of fundus images in various color spaces.
For better evaluation, we score the quality of the enhanced image as an indicator of high-quality images, denoted as $s_{HQ}$.
Fig.~\ref{fig:fig3}b and d display the $s_{HQ}$ values for images enhanced by various methods. 
UniMIE performs at the best level among all image enhancement methods. 
These results demonstrate that thanks to the inherited powerful prior, UniMIE exhibits excellent zero-shot capabilities.

\subsection{Evaluation over Endoscene images}
An endoscope is a common tool used to examine the internal organs and tissues of the human body, such as gastroscopy, colonoscopy, and bronchoscopy. However, due to factors such as natural damping and light scattering of tissues, endoscopic images often suffer from problems including low contrast, blur, and noise, which limit doctors' observation and diagnosis of lesions~\cite{goetz2014microscopic}. Through endoscopic image enhancement technology, the clarity, contrast and details of the image can be improved, allowing doctors to detect and diagnose lesions more accurately, and improve the accuracy and reliability of diagnosis.

Therefore, we conducted a comparison of the light distribution on the Endoscopy dataset (Fig.~\ref{fig:fig3}e, Supplementary Fig. 4). 
It is evident that the original image exhibits uneven light distribution. 
Supervised learning and semi-supervised image enhancement methods have demonstrated the ability to enhance light intensity. However, these approaches frequently suffer from overexposure, thereby compromising the preservation of original texture details.
Conversely, unsupervised methods, like Zero-DCE and Zero-DCE++, often introduce modifications to the color tones of medical images. Consequently, the enhanced images produced by these methods may deviate from the authentic appearance of high-quality real-world medical images.
In contrast, UniMIE yields the most favorable visual outcomes, generating enhanced images that closely resemble the illumination distribution of high-quality medical images in the real world.

Four image quality assessment metrics were employed to quantitatively evaluate the enhanced medical images using the proposed UniMIE and other methods on the Endoscene dataset, including the Lightness Order Error (LOE)~\cite{wang2013naturalness}, blind/reference-free image spatial quality evaluator (BRISQUE)~\cite{mittal2012making}, perception-based image quality evaluator (PIQE)~\cite{venkatanath2015blind}, and Entropy~\cite{tsai2008information}. 
The smaller the first three metrics the better, the larger the last the better.
Fig.~\ref{fig:fig3}f-i illustrates the results obtained from employing various techniques for endoscopic image enhancement. It is evident that among the zero-shot learning methods, UniMIE excels in terms of LOE, BRISQUE, PIQE, and Entropy metrics. 

\begin{figure}[t]
    \centering
    \includegraphics[width=0.95\linewidth]{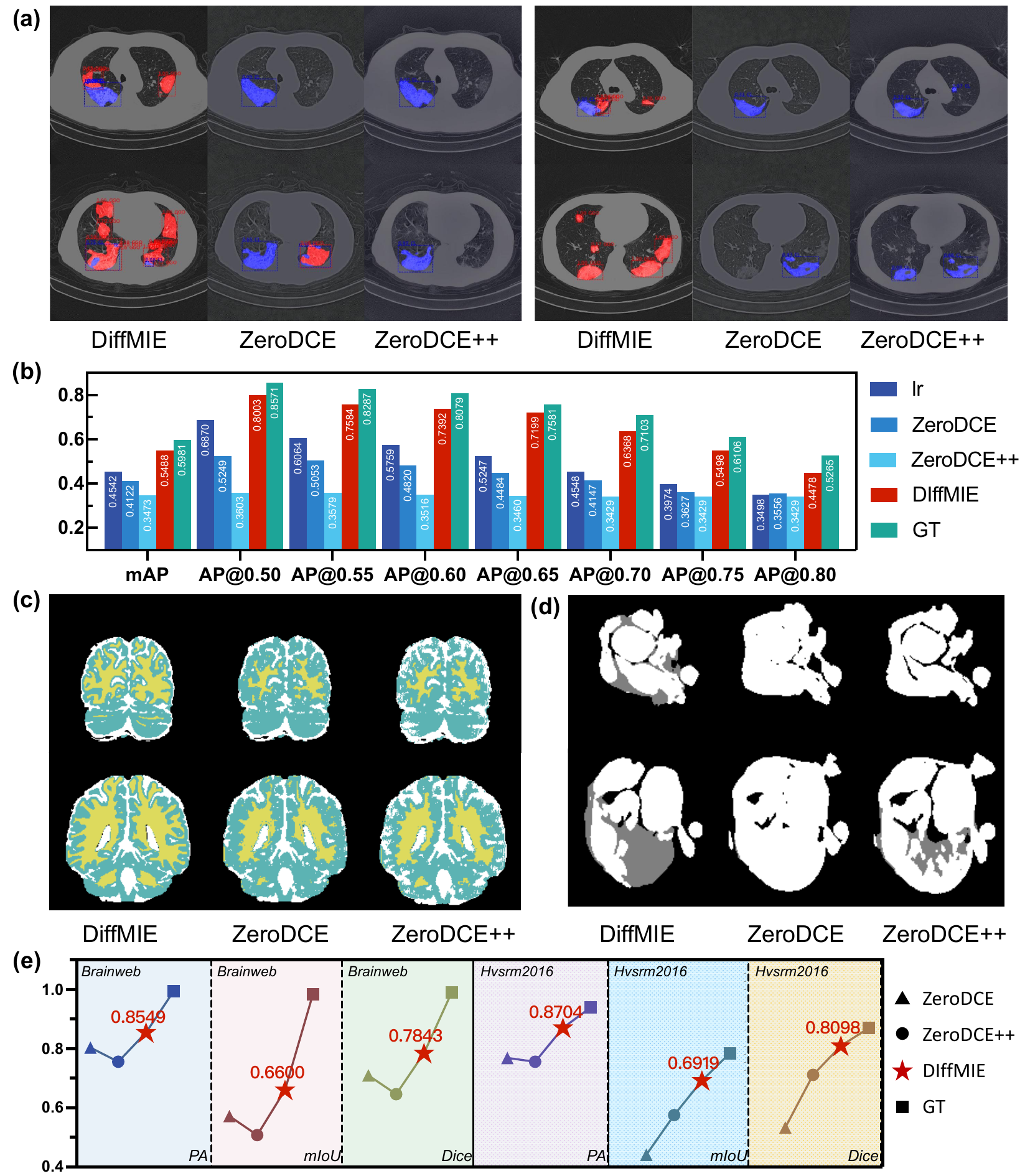}
    \caption{\textbf{a,} The detection results of COVID-19 after enhancement. \textbf{b,} Average precision of segmentation results on enhanced medical images. \textbf{c,} Segmentation results on enhanced images from BrainWeb and HVSMR. \textbf{d,} Comparisons of PA, JS, and Dice metrics on BrainWeb and HVSMR.}
    \label{fig:fig4}
\end{figure}

\subsection{Transfering UniMIE to CT and MRI images}
CT (Computed Tomography) and MRI (Magnetic Resonance Imaging) images can be influenced by various factors, including noise, artifacts, low contrast, and other similar issues. Medical image enhancement technology improves image quality by reducing noise levels, reducing artifacts, and increasing image contrast and clarity. CT and MRI image enhancement can highlight lesions (such as tumors, injuries, abnormal tissue, etc.), making it easier for doctors to detect and locate these abnormal areas. This is critical for early detection of lesions, correct diagnosis and treatment planning~\cite{leveridge2010imaging}.

To demonstrate the versatility of UniMIE towards other modality images, we further evaluate UniMIE on CT and MRI.
Given the inherent dissimilarities between the modalities of CT and MRI images and the training set of other methods, we have opted to conduct a comparative analysis of the Zero-DCE and Zero-DCE++ techniques, both of which fall under the category of zero-shot approaches.
Compared with Zero-DCE and Zero-DCE++, UniMIE can also generalize well to CT and MRI image enhancement (Supplementary Fig. 5-7).
We further conduct COVID-19 detection experiments on the enhanced CT images.
The diagnostic utility of characteristic lesions, such as Ground Glass Opacity (GGO) and Consolidation (C), in chest CT scans has been demonstrated for diagnosing COVID-19 and common pneumonia (CP)~\cite{ter2022covid}.
In COVID-19 patients, GGO is more frequently observed and tends to be bilateral. 
Additionally, subsegmental C areas are more prevalent in COVID-19 patients compared to those with CP. 
The majority of COVID-19 patients display either GGO, Consolidation or a combination of both. 
Therefore, the enhanced images will help doctors make more precise decisions, while the low-quality conditions will exert negative influences on the recognition of GGO and C.
Following COVID-19 detection~\cite{ter2020detection,ter2020lightweight}, we conduct down-stream experiments.
As shown in Fig.~\ref{fig:fig4}a, more GGO and C areas are detected on the enhanced images by our UniMIE.
In comparison, the detection method can not perform well on the images enhanced by Zero-DCE and Zero-DCE++.
These results can be further validated by quantitative comparison (Fig.~\ref{fig:fig4}b).
The average detection precision of UniMIE approaches the upbound of the detection model.

Moreover, we also perform enhancement experiments on MRI datasets (BrainWeb~\cite{Cocosco1997BrainWebOI} and HVSMR~\cite{Pace2015InteractiveWS}) (Supplementary Figs. 6 and 7).
After enhancement, we follow RDC~\cite{wen2020segmenting} to utilize the Recurrent Decoding Cell to conduct segmentation experiments.
The images enhanced by UniMIE can obtain high-detailed segmentation results both on BrainWeb and HVSMR (Fig.~\ref{fig:fig4}c and d), compared with Zero-DCE and Zero-DCE++.
The quantitative comparison in Fig.~\ref{fig:fig4}e, Pixel Accuracy (PA), Dice and mean Intersection over Union (mIoU) metrics also demonstrate that the images enhanced by UniMIE can maintain the details of medical images for better segmentation.

\begin{figure}[t]
    \centering
    \includegraphics[width=\linewidth]{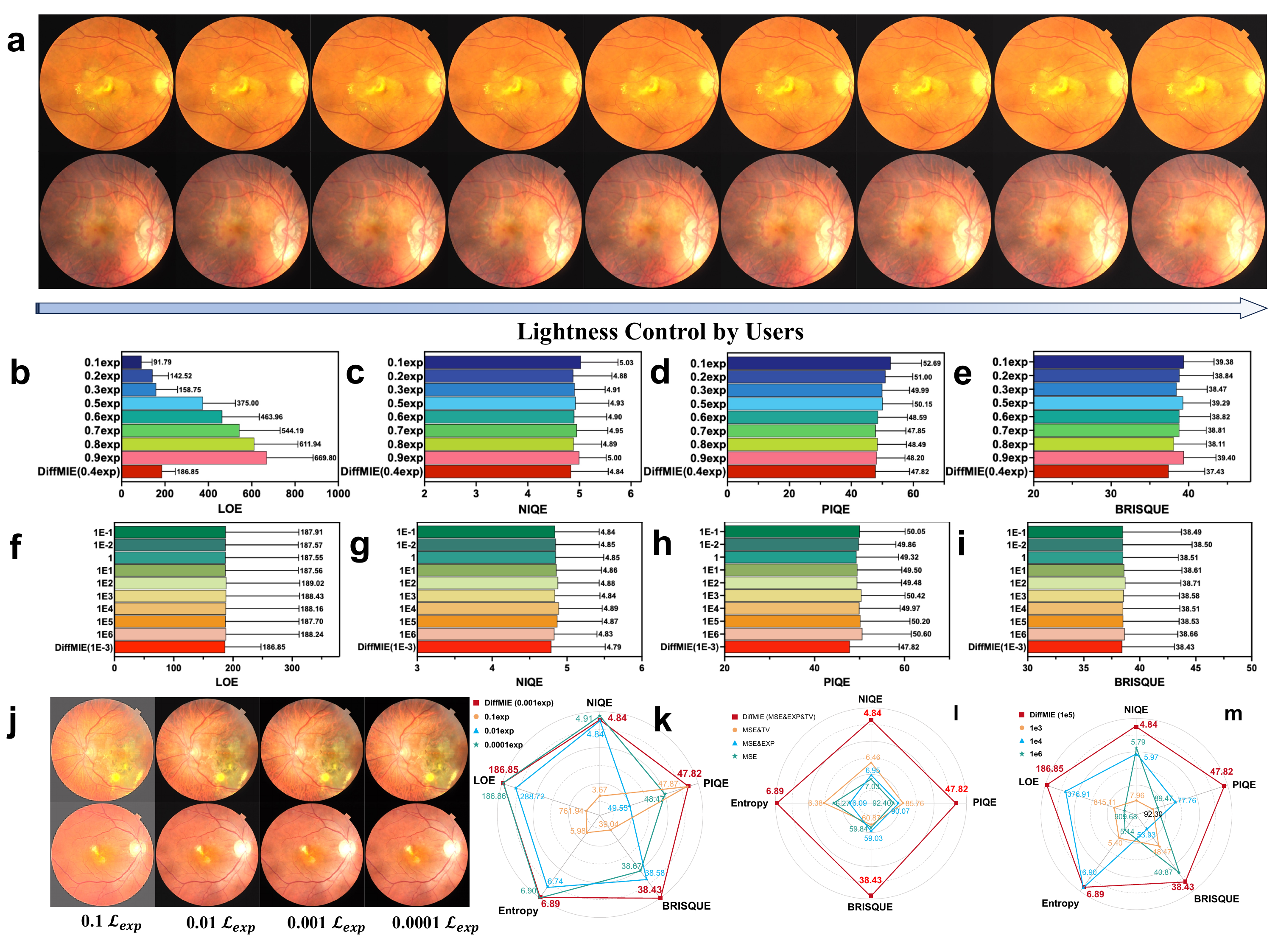}
    \caption{\textbf{a,} Visualization of lightness control. Performance of various well-exposedness levels: \textbf{b,} LOE, \textbf{c,} NIQE, \textbf{d,} PIQE, and \textbf{e,} BRISQUE. Performance of various weights of illumination smoothness loss: \textbf{f,} LOE, \textbf{g,} NIQE, \textbf{h,} PIQE, and \textbf{i,} BRISQUE. \textbf{j,} Visualization and \textbf{k,} performance of different weights of exposure loss. \textbf{l,} Ablation studies on the combination of losses. \textbf{m,} Ablation studies on the guidance scale.}
    \label{fig:fig5}
\end{figure}

\subsection{Adjustable well-exposedness level.}
Medical images often exhibit variations in brightness levels and contrast. By modifying the image's brightness, medical professionals and observers can effectively adapt to the specific display environment and image requirements, ultimately enhancing the observation experience.
Appropriate brightness adjustment can improve the contrast of the image and make the boundaries between different tissues or structures more clearly visible. This helps doctors more accurately identify and locate lesions and make more precise diagnoses.

The incorporation of the exposure control loss allows for effective regulation of the level of well-exposedness in the enhanced images. Fig.~\ref{fig:fig5}c illustrates that the adjustable well-exposedness level of the generated images increases as the well-exposedness level ($E$) is increased. This quantitative trend is further supported by Fig.~\ref{fig:fig5}d-g, which demonstrates that our UniMIE model with a well-exposedness level of 0.4 achieves the best NIQE and PIQE scores, as well as the second-best BRISQUE score. It is worth noting that the LOE metric, which assesses the light order of images, favors darker images in terms of achieving better scores.

\section{Discussion}
\label{sec:discussion}

This paper induces UniMIE, a unified solver designed to enhance medical images of various anatomical structures and lesions across different medical imaging modalities. (Supplementary Fig. 1-15). 
Unlike previous approaches, UniMIE does not depend on a training procedure. Instead, it leverages a pre-trained model on ImageNet to demonstrate its unsupervised enhancement capabilities on various medical image datasets.
Its capabilities to operate on any-sized images and adjustable lightness according to the necessity of doctors, make UniMIE a versatile tool for universal medical image enhancement.

The effectiveness of UniMIE has been extensively evaluated on 15 distinct medical image datasets with 13 medical image modalities. 
The experimental results highlight the superior performance of UniMIE compared to other low-light enhancement methods. 
Furthermore, we have undertaken a comprehensive analysis to assess the influence of our UniMIE on several medical image analysis and clinical tasks, including tortuosity grading, nerve segmentation, COVID-19 segmentation, and so on.
Given that UniMIE is a training-free method, we plan to integrate it into existing medical equipment, including microscopes and gastroenteroscopes, to further evaluate and validate its practical application. 
This integration will allow us to assess the effectiveness of UniMIE in real-world medical scenarios and gauge its potential for enhancing medical image quality directly within the existing infrastructure.

UniMIE does not require any assumptions regarding the source, type, individual patient differences, or medical device variations of medical images. It can be applied to any medical image without the need for training. This remarkable generalization capability further confirms the significant clinical value of UniMIE. For instance, during laparoscopic surgery, there is no need for additional light sources to ensure image brightness. The enhanced images can greatly assist doctors in performing more precise treatments. Currently, the main drawback of UniMIE is the increase in enhancement time as image resolution increases. This is primarily due to our unconditional diffusion model's resolution being limited to 256 $\times$ 256, causing our devised patch-based method to scale with the image resolution. As resolution increases, the number of patches also increases, resulting in longer enhancement times. However, UniMIE serves as a method framework and can be further developed with the advancement of artificial intelligence. Higher resolution and faster diffusion models can propel the progress of UniMIE.

To conclude, this paper emphasizes the viability of developing a unified pre-trained model that can effectively handle various enhancement tasks, thus obviating the necessity for task-specific models. UniMIE, as a training-free diffusion model utilized in medical image enhancement, exhibits significant potential for expediting the development of novel diagnostic and therapeutic tools, ultimately leading to enhanced patient care.

\section{Methods}

\subsection{Denoising diffusion probabilistic models}

Denoising diffusion probabilistic models~\cite{guo2024diffusion,huang2024dual} are capable of transforming a complex data distribution $\mathbf{x}_0 \sim p_{\text {data}}$ into a simpler latent distribution $\mathbf{x}_T \sim p_{\text {latent }}=\mathcal{N}\left(\mathbf{0}, \boldsymbol{I}\right)$, where $\mathcal{N}$ represents the Gaussian distribution. Subsequently, the data can be recovered from the noise distribution. Denoising diffusion probabilistic models mainly consist of the forward diffusion process and the reverse denoising process.

\noindent\textit{\underline{The forward diffusion process}} repeatedly introduces noise to initial clean data $\mathbf{x}_0$, gradually converging towards Gaussian noise $p_{\text {latent }}$ at forward diffusion time steps $T$. Noisy data $\mathbf{x}_1, \ldots, \mathbf{x}_T$ are sampled from data $p_{\text {data}}$ using a forward diffusion process, characterized by a Gaussian transition:
\begin{equation}
     \!\!q\left(\mathbf{x}_1, \cdots, \mathbf{x}_T \mid \mathbf{x}_0\right)=\prod_{t=1}^T q\left(\mathbf{x}_t \mid \mathbf{x}_{t-1}\right), 
\end{equation}
where $t$ represents for forward step, $q\left(\mathbf{x}_t \mid \mathbf{x}_{t-1}\right)=\mathcal{N}\left(\mathbf{x}_t; \sqrt{1-\beta_t} \mathbf{x}_{t-1}, \beta_t \boldsymbol{I}\right)$, while $\beta_t$ is fixed or learnable variance schedule. 
One crucial characteristic of the forward diffusion process is its ability to directly sample any step $\mathbf{x}_t$ from $\mathbf{x}_0$ using:
\begin{equation}
\mathbf{x}_t=\sqrt{\bar{\alpha}_t} \mathbf{x}_0+\sqrt{1-\bar{\alpha}_t} \boldsymbol{\epsilon}, 
\label{eq2}
\end{equation}
where $\boldsymbol{\epsilon} \sim \mathcal{N}(0, \boldsymbol{I})$, $\alpha_t=1-\beta_t$ and $\bar{\alpha}_t=\prod_{i=1}^t \alpha_i$. 
Therefore, the closed-form expression is $q\left(\mathbf{x}_t \mid \mathbf{x}_0\right)$
Herein, $\bar{\alpha}_t$ tends to approach zero as the value of $T$ increases, while the distribution $q\left(\mathbf{x}_T \mid \mathbf{x}_0\right)$ approximates the noise distribution $p_{\text{latent}}$.

\noindent\textit{\underline{The reverse denoising process}} gradually denoises a Gaussian noise to generate clean data. The process begins with a noise vector $\mathbf{x}_T$ sampled from a Gaussian distribution. 
The conversion from $\mathbf{x}_T$ to $\mathbf{x}_0$ is formally formulated as follows:
\begin{equation}
\begin{split}
& p_{\boldsymbol{\theta}}\left(\mathbf{x}_0, \cdots, \mathbf{x}_{T-1} \mid \mathbf{x}_T\right)=\prod_{t=1}^T p_{\boldsymbol{\theta}}\left(\mathbf{x}_{t-1} \mid \mathbf{x}_t\right),\quad \\ &
p_{\boldsymbol{\theta}}\left(\mathbf{x}_{t-1} \mid \mathbf{x}_t\right)=\mathcal{N}\left(\mathbf{x}_{t-1} ; \boldsymbol{\mu}_{\boldsymbol{\theta}}\left(\mathbf{x}_t, t\right), \Sigma_{\theta} \boldsymbol{I}\right).
\end{split}
\label{eq:reverse}
\end{equation}
The objective is to estimate the mean $\boldsymbol{\mu}_{\boldsymbol{\theta}}\left(\mathbf{x}_t, t\right)$ using a neural network $\boldsymbol{\theta}$~\cite{ho2020denoising}. 
Furthermore, the symbol $\Sigma_{\theta}$ can take on two forms: learnable parameters~\cite{nichol2021improved}, or time-dependent constants~\cite{ho2020denoising}.
The function approximator $\epsilon_{\theta}$ is designed to predict $\epsilon$ based on the input $\mathbf{x}_t$, as follows:
\begin{equation}
    \boldsymbol{\mu}_\theta\left(\mathbf{x}_t, t\right)=\frac{1}{\sqrt{\alpha_t}}\left(\mathbf{x}_t-\frac{\beta_t}{\sqrt{1-\bar{\alpha}_t}} \boldsymbol{\epsilon}_\theta\left(\mathbf{x}_t, t\right)\right)
\end{equation}

In practical applications, it is customary to predict $\mathbf{\hat{x}}_0$ based on the given input $\mathbf{x}_t$.
Subsequently, the sampling of $\mathbf{x}_{t-1}$ is performed by using $\mathbf{\hat{x}}_0$ and $\mathbf{x}_t$:
\begin{equation}
\!\!\!\!  \mathbf{\hat{x}}_{0} =  \frac{\mathbf{x}_{t}}{\sqrt{\bar{\alpha}_{t}}}-\frac{\sqrt{1-\bar{\alpha}_{t}} \epsilon_{\theta}\left(\mathbf{x}_{t}, t\right)}{\sqrt{\bar{\alpha}_{t}}}
\label{eq5}
\end{equation}
\begin{equation}
\begin{split}
\!\!\!\!  q\left(\mathbf{x}_{t-1} \mid \mathbf{x}_t, \mathbf{\hat{x}}_0\right) & =\mathcal{N}\left(\mathbf{x}_{t-1} ; \tilde{\mathbf{\mu}}_t\left(\mathbf{x}_t, \mathbf{\hat{x}}_0\right), \tilde{\beta}_t \mathbf{I}\right), \\
\!\!\!\!  \text { where } \quad \tilde{\boldsymbol{\mu}}_t\left(\mathbf{x}_t, \mathbf{\hat{x}}_0\right) & =\frac{\sqrt{\bar{\alpha}_{t-1}} \beta_t}{1-\bar{\alpha}_t} \mathbf{\hat{x}}_0+\frac{\sqrt{\alpha_t}\left(1-\bar{\alpha}_{t-1}\right)}{1-\bar{\alpha}_t} \mathbf{x}_t \quad \\
\!\!\!\!  & \text { and } \quad \tilde{\beta}_t=\frac{1-\bar{\alpha}_{t-1}}{1-\bar{\alpha}_t} \beta_t
\end{split}
\label{eq4}
\end{equation}

\subsection{Conditional diffusion models}



Our objective is to utilize a pre-trained DDPM as a powerful prior for universal medical image enhancement, specifically when dealing with low-quality medical images of various modalities.
We assume a universal scenario where a degraded medical image, denoted by $\mathbf{y}$, is obtained through the process of capturing the original high-quality medical image, represented by $\mathbf{x}$, using a degradation model denoted by $\mathcal{D}$.
Considering $\mathbf{y}$ as the deteriorated observations of $\mathbf{x}$, we leverage the statistical information of $\mathbf{x}$ stored in prior knowledge. We then search within the space of $\mathbf{x}$ to find an optimal alignment with $\mathbf{y}$.
The objective of this study is to examine a broader and more generalized image prior, focusing specifically on diffusion models that have undergone pre-training using extensive collections of natural images~\cite{dhariwal2021diffusion}.
The reverse denoising process could be conditioned on the $\mathbf{y}$~\cite{saharia2022image, rombach2022high, choi2022perception}. 
In detail, the reverse denoising distribution $p_{\boldsymbol{\theta}}(\mathbf{x}_{t-1}|\mathbf{x}_{t})$ in equation (\ref{eq:reverse}) is transformed as a conditional distribution $p_{\mathbf{\theta}}\left(\mathbf{x}_{t-1} |\mathbf{x}_t, \mathbf{y}\right)$:
\begin{equation}
\begin{split}
    \!\!\!\! \log p_{\boldsymbol{\theta}}\left(\mathbf{x}_{t-1}|\mathbf{x}_t, \mathbf{y}\right) &=\log \left(p_{\boldsymbol{\theta}}\left(\mathbf{x}_{t-1}|\mathbf{x}_t\right) p\left(\mathbf{y}| \mathbf{x}_t\right)\right)+Z_1 
    \\ \!\!\!\!  & \approx \log p(\boldsymbol{r})+Z_2, 
\end{split}
\end{equation}
where $\boldsymbol{r} \sim \mathcal{N}\left(\boldsymbol{r} ; \boldsymbol{\mu}_{\boldsymbol{\theta}}\left(\mathbf{x}_t, t\right)+\Sigma \boldsymbol{g}, \Sigma \right)$ and $\boldsymbol{g}=\nabla_{\mathbf{x}_t} \log p\left(\mathbf{y} \mid \mathbf{x}_t\right)$.
To maintain conciseness, we denote that $\Sigma = \Sigma_{\theta}\left(\mathbf{x}_{t}\right)$. 
$Z_1$ and $Z_2$ are constants, $p_{\boldsymbol{\theta}}\left(\mathbf{x}_{t-1} \mid \mathbf{x}_t\right)$ is determined by equation (\ref{eq:reverse}). 
$p(\mathbf{y}|\mathbf{x}_t)$ can be interpreted as the probability of $\mathbf{x}_t$ being denoised into a high-quality image that matches $\mathbf{y}$, where an approximation is formulated:
\begin{equation}
\begin{split}
    p\left(\mathbf{y} \mid \mathbf{x}_t\right) = \frac{1}{K} & \exp \left(-\left[s\mathcal{L}\left(\mathcal{D}(\mathbf{x}_t), \mathbf{y}\right) + \lambda \mathcal{Q}(\mathbf{x}_t) \right]\right),
\end{split}
\end{equation}
where $\mathcal{L}$ represents the Mean Square Error (MSE), $Z$ denotes a normalization factor, and $s$ serves as a scaling factor that regulates the magnitude of guidance.
This formulation promotes consistency between $\mathbf{x}_t$ and the corrupted medical image $\mathbf{y}$ in order to maximize the probability $p\left(\mathbf{y} \mid \mathbf{x}_t\right)$.
$\mathcal{Q}$ represents an optional loss for quality enhancement (refer to the Loss function), aiming to increase the flexibility of UniMIE. It enables the control of the well-exposeness level and enhances the overall quality of generated medical images.
The scale factor $\lambda$ is used to adjust the quality of images. 
The computation of gradients on both sides is performed in the following manner:
\begin{equation}
    \begin{aligned}
    \!\!\!\! &\log p\left(\mathbf{y} \mid \mathbf{x}_t\right)=-\log Z-s \mathcal{L}\left(\mathcal{D}(\mathbf{x}_t), \mathbf{y}\right) - \lambda \mathcal{Q}\left(\mathbf{x}_t\right) \\
    \!\!\!\! &\nabla_{\mathbf{x}_t} \log p\left(\mathbf{y} \mid \mathbf{x}_t\right)=-s \nabla_{\mathbf{x}_t} \mathcal{L}\left(\mathcal{D}(\mathbf{x}_t), \mathbf{y}\right) - \lambda \nabla_{\mathbf{x}_t} \mathcal{Q}\left(\mathbf{x}_t\right),
    \end{aligned}
\end{equation}
where the conditional transition $p_{\boldsymbol{\theta}}\left(\mathbf{x}_{t-1} \mid \mathbf{x}_t, \mathbf{y}\right)$ can be approximately acquired through the unconditional transition $p_{\boldsymbol{\theta}}\left(\mathbf{x}_{t-1} \mid \mathbf{x}_t\right)$ by the mean shifting of the unconditional distribution by $- (s \Sigma \nabla_{\mathbf{x}_t} \mathcal{L}\left(\mathcal{D}(\mathbf{x}_t), \mathbf{y}\right) + \lambda \Sigma \nabla_{\mathbf{x}_t} \mathcal{Q}(\mathbf{x}_t))$.

In this study, the conditional signal is applied on $\mathbf{\hat{x}}_0$. During the enhancement process, the pre-trained DDPM typically begins by predicting a clean medical image $\mathbf{\hat{x}}_0$ from the noisy medical image $\mathbf{x}_{t}$ (Supplementary Algorithm 1). This prediction is achieved by estimating the noise present in $\mathbf{x}_{t}$, directly inferred from equation (\ref{eq5}) at each $t$.
Subsequently, the predicted $\mathbf{\hat{x}}_0$ and the current noisy medical image $\mathbf{x}_{t}$ are employed to sample the latent variable $\mathbf{x}_{t-1}$ for the next step.
To exert control over the generation process of the DDPM, guidance can be incorporated for the intermediate variable $\mathbf{\hat{x}}_0$. 

\subsection{Degradation function for medical image enhancement}
In the field of clinical medicine, numerous images undergo intricate degradation processes due to the complex nature of the human body and the limitations of signal acquisition equipment~\cite{zhang2021designing}. These degradation processes often involve unknown degradation functions or parameters~\cite{wang2021real}. 
In this particular scenario, it is necessary to estimate both the high-quality images and the unknown parameters of the degradation functions simultaneously.
In this study, the task of medical image enhancement can be viewed as dealing with unknown degradation functions. To address this, we propose a straightforward yet efficient degradation model that effectively simulates the complex degradation. The formulation of this function:
\begin{equation}
    \mathbf{y} = f \mathbf{x}+\boldsymbol{\mathcal{M}}, 
\label{eq:light}
\end{equation}
where the enhancement factor, denoted as scalar $f$, and the enhancement mask, represented by the vector $\boldsymbol{\mathcal{M}}$ that shares the same dimension with $\mathbf{x}$, are both unknown parameters of the degradation function.
The rationale for utilizing this straightforward degradation model stems from the fact that the transformation between any set of distorted images and their corresponding high-quality image can be encapsulated by the variables $f$ and $\boldsymbol{\mathcal{M}}$.
If the sizes of $\mathbf{y}$ and $\mathbf{x}$ do not match, we can initially adjust the size of $\mathbf{y}$ to match that of $\mathbf{x}$ before applying this degradation function.
In order to estimate both $f$ and $\boldsymbol{\mathcal{M}}$ for each degraded image, a random initialization is performed. This initialization is then followed by a synchronous optimization process, which operates in reverse to the DDPMs process. Supplementary Algorithm 1 provides a detailed representation of this procedure.

\subsection{Any-size medical image enhancement with patch-based method}

Medical image enhancement datasets consist of images with varying dimensions. However, the existing generative architectures are predominantly designed to process images of fixed sizes. 
Our approach entails the decomposition of images into overlapping fixed-sized patches during the testing phase, which are subsequently merged during the sampling process. This method distinguishes itself from a simplistic baseline approach that involves averaging overlapping final reconstructions after sampling. Employing such a method after sampling may compromise the fidelity of the local patch distribution to the learned posterior~\cite{whang2022deblurring}.

The fundamental principle that drives patch-based restoration is the execution of localized operations on image patches, followed by the efficient merging of the resultant outcomes.
Indeed, the independent restoration of intermediate patches in patch-based restoration has posed a significant drawback, leading to the emergence of merging artifacts in the resulting image.
This issue has received considerable attention in traditional restoration methods. To address this challenge, we propose to integrate a guided reverse sampling process. This process ensures consistency among neighboring patches, mitigating the issue of merging artifacts.

Specifically, the unknown ground truth with arbitrary size is defined as $\mathbf{X}_0$ and the low-quality observation is denoted as $\mathbf{Y}$. $\boldsymbol{P}_i$ represents a binary mask matrix. 
Both $\mathbf{X}_0$ and $\mathbf{Y}$ have the same dimensions, while $\boldsymbol{P}_i$ indicates the location of the $i$-th patch, which is of size $p \times p$, within the image.
The conditional reverse process can be formulated as:
\begin{equation}
   p_\theta\left(\mathbf{x}_{0: T}^{(i)} \mid \mathbf{y}^{(i)}\right)=p\left(\mathbf{x}_T^{(i)}\right) \prod_{t=1}^T p_\theta\left(\mathbf{x}_{t-1}^{(i)} \mid \mathbf{x}_t^{(i)}, \mathbf{y}^{(i)}\right).
\end{equation}
By utilizing the Crop(.) operation, we can extract $p \times p$ patches from the unknown ground truth image $\mathbf{X}_0$ and the low-quality observation $\mathbf{Y}$. Specifically, $\mathbf{x}_0^{(i)}$ represents the patch obtained by cropping the element-wise product of $\boldsymbol{P}_i$ and $\mathbf{X}_0$, while $\mathbf{y}^{(i)}$ represents the patch obtained by cropping the element-wise product of $\boldsymbol{P}_i$ and $\mathbf{Y}$.


The details of our patch-based approach for enhancing medical images are depicted in Figure~\ref{fig:fig1} and described in Supplementary Algorithm 2.
Initially, the medical image $\mathbf{Y}$ of arbitrary dimensions undergoes decomposition through the extraction of overlapping $p \times p$ patches. This process is implemented using a grid-like parsing scheme.
Throughout the entire image, a grid-like arrangement is utilized, wherein each grid cell is composed of $r \times r$ pixels, with $r$ being smaller than $p$.
By traversing this grid using a step size of $r$ in both the horizontal and vertical dimensions, the extraction of all $p \times p$ patches is achieved.
$N$ is defined as the total number of patches.
Then a dictionary is established that records the locations of these overlapping patches.

Given the inherently ill-posed nature of the problem, it is important to consider that the utilization of neighboring overlapping patches for conditional reverse sampling may result in varying restoration estimates for grid cells.
We mitigate this problem by conducting reverse sampling that relies on the average estimated noise for each pixel within the overlapping regions at every time step $t$ (see Figure~\ref{fig:fig1}).
Our proposed method guides the reverse denoising process, thereby ensuring enhanced fidelity across all neighboring patches.
At each denoising time step $t$ of the sampling process, the following steps are performed: 
(1) The function $\boldsymbol{\epsilon}_\theta\left(\mathbf{x}_t^{(n)}, \mathbf{y}^{(n)}, t\right)$ is leveraged to estimate the additive noise for all overlapping patches $n \in \{1, \ldots, N\}$.
(2) These overlapping noise estimations are accumulated at their respective patches in a matrix $\hat{\Theta}_t$, which has the same size as the entire image. 
(3) The noise estimates that overlap are accumulated in a matrix $\hat{\Theta}_t$ at their respective patch locations. This matrix has the same dimensions as the entire image.
(4) An implicit sampling update is conducted utilizing the smoothed estimate, denoted as $\hat{\Theta}_t$, of the noise present throughout the entire image.

It is noteworthy that a smaller value of $r$ results in increased overlap between patches, leading to smoother outcomes. However, this also entails a higher computational load. In our study, we employed a patch size of $p=256$ pixels for $\boldsymbol{P}_i$, along with a patch overlap of $r=128$ pixels. Before processing, we resized the dimensions of the entire image to ensure they were multiples of 16.
Hence, selecting $r=p$ would result in a collection of non-overlapping patches, assuming independence among the patches during the enhancement process. However, it is important to note that neighboring patches in images are not independent. Therefore, such an approach would yield a suboptimal approximation, introducing edge artifacts in the enhanced images.

\subsection{Loss function}
\label{sec:loss}

We utilize Mean Square Error as distance metric $\mathcal{L}$.
The quality enhancement losses $\mathcal{Q}$ encompass both the exposure control loss as well as the illumination smoothness loss.

\textbf{Exposure Control.}
In order to mitigate the issue of under- or over-exposed areas, we have developed an exposure control loss, denoted as $\boldsymbol{\mathcal{L}}_{\text{EC}}$, which serves to regulate the exposure level.
In practice, this exposure control loss quantifies the disparity between the average intensity of local regions and the desired level of well-exposedness $E$.
To establish the value of $E$, we adopt the convention outlined in previous works~\cite{guo2020zero}, which sets $E$ as the gray level within the RGB color space. Based on ablation studies conducted in our experiments, we set $E$ to 0.4. The expression for the exposure control loss $\boldsymbol{\mathcal{L}}_{\text{EC}}$ is given by:
\begin{equation}
    \boldsymbol{\mathcal{L}}_{\text{ EC}}=\frac{1}{O} \sum_{k=1}^O\left|R_k-E\right|,
\end{equation}
where $O$ denotes the total count of non-overlapping local regions with dimensions of $16 \times 16$. 
Meanwhile, $R$ represents the average intensity of local regions within the enhanced medical image.

\textbf{Illumination Smoothness.}
To maintain the desired monotonicity relations between adjacent pixels, we integrate an illumination smoothness loss term $\boldsymbol{\mathcal{L}}_{is_{\mathcal{C}}}$ into each curve parameter map $\mathcal{U}$.
$\boldsymbol{\mathcal{L}}_{is_{\mathcal{C}}}$ is defined as follows:
\begin{equation}
    \boldsymbol{\mathcal{L}}_{i s_{\mathcal{U}}}=\frac{1}{M} \sum_{n=1}^M \sum_{c \in \xi}(\left|\nabla_x \mathcal{U}_n^c\right|+\left|\nabla_y \mathcal{U}_n^c \mid\right)^2, \xi=\{R, G, B\} \text {, }
\end{equation}
where $M$ represents the number of iterations, while $\nabla_x$ and $\nabla_y$ denote the horizontal and vertical gradient operations, respectively.

\subsection{Effectiveness on the weight of illumination smoothness loss.} 
To validate the appropriate weight for the illumination smoothness loss, additional experiments were conducted. Fig.~\ref{fig:fig5}f-i reveals that when the weight is set to 0.001, the UniMIE model achieves the highest scores for LOE, PIQE, and BRISQUE. This indicates that assigning this weight value optimizes the performance of the model in terms of these evaluation metrics.

\subsection{Effectiveness on the weight of exposure control loss.}
Additional experiments were undertaken to investigate the impact of varying the weight of the exposure control loss. 
Fig.~\ref{fig:fig5}j visually demonstrates that a significant weight assigned to the exposure control loss can result in over-exposed images.
The quantitive results, as presented in Fig.~\ref{fig:fig5}k, highlight that the UniMIE with an exposure control loss weight of 0.001 demonstrated superior performance in terms of LOE, PIQE, and BRISQUE metrics.

\subsection{Effectiveness on the combination of losses.} 
Firstly, we further carried out ablation studies to examine the effects of integrating various loss functions. The findings, as depicted in Fig.~\ref{fig:fig5}n, reveal that our UniMIE, when supplemented with exposure control loss and illumination smoothness loss, yielded the most favorable outcomes in terms of Entropy, NIQE, PIQE, and BRISQUE. These results serve as compelling evidence for the effectiveness of our approach in leveraging these quality enhancement losses.

\subsection{Effectiveness on the guidance scale.}
The guidance scale, being the most significant super-parameter of our UniMIE model, holds significant influence over the quality of the generated images. Fig.~\ref{fig:fig5}m demonstrates that setting the guidance scale to 100,000 yields the best performance, thereby highlighting the importance of carefully selecting this parameter to optimize the overall results.

\subsection{Statistical analysis}
To assess the quantitative performance of the medical image enhancement results, we employed standard image quality assessment metrics, as described below:

PIQE~\cite{mittal2012making}, BRISQUE~\cite{mittal2012no}, and Entropy~\cite{tsai2008information} are utilized to measure the non-reference enhanced image qualities.

The metric of LOE is employed to objectively measure the degree of lightness distortion in the enhanced results. LOE is defined as:
\begin{equation}
    L O E=\frac{1}{m} \sum_{k=1}^m ROD(k),
\end{equation}
where $ROD(x)$ denotes the relative order difference of the lightness between the input image and its enhanced version for a given pixel $k$. This metric is defined as follows:
\begin{equation}
    ROD(k)=\sum_{g=1}^m T(\mathbf{H}(k), \mathbf{H}(y)) \oplus U\left(\mathbf{H}^{\prime}(k), \mathbf{H}^{\prime}(g)\right),
\end{equation}
$m$ represents the pixel number, $\oplus$ symbolizes the exclusive-or operator, $\mathbf{H}(k)$ and $\mathbf{H}^{\prime}(k)$ correspond to the maximum values among the three channels at location $k$ in the original images and the enhanced images. The function $T(p, q)$ returns a value of 1 if $p$ is greater than or equal to $q$, and 0 otherwise.

Moreover, we use PA, mIoU, and Dice metrics to measure the segmentation results after enhancement.

\subsection{Software utilized}
All the enhancement codes were implemented using Python (3.9.15) and PyTorch (1.12.1) as the chosen deep-learning framework.
Moreover, for data analysis and visualization, several Python packages were utilized, including torchvision (0.13.1), numpy (1.23.5), scikit-image (0.20.0), scipy (1.9.1), pandas (2.0.0), matplotlib (3.5.2), opencv-python (4.6.0), and plotly (5.14.1).
EdrawMax was used to create Fig.~\ref{fig:teaser}.

\section{Data availability}
The datasets utilized for validation in this work are accessible in the public domain and can be obtained by following the provided links at \url{https://github.com/Fayeben/UniMIE}.
We have duly confirmed that all the image datasets employed in this work are publicly accessible and explicitly authorized for research purposes.

\section{Code availability}
The inference scripts are publicly available at \url{https://github.com/Fayeben/UniMIE}. And the pre-trained Diffusion Model can be found at \url{https://github.com/openai/guided-diffusion}.


\bibliographystyle{naturemag_doi} 
\bibliography{ref.bib}

\setstretch{1}

\end{document}